\documentclass[onecolumn,preprintnumbers,prd,showpacs,amsmath,amssymb]{revtex4}
\usepackage[dvips]{graphicx}
\usepackage{amsthm}
\usepackage{amscd}
\usepackage{amstext}
\usepackage{amsfonts}
\usepackage{hyperref}
\newcommand{\p}{\partial}

\newcommand{\dd}{{\rm d}}

\begin{document}

\title[A globally well-behaved simultaneity connection]{A globally well-behaved simultaneity connection for stationary frames \\ in the weak field limit} 

\thanks{Published version: E. Minguzzi, Class. Quantum Grav. 21 (2004)
4123-4146. Classical and Quantum Gravity copyright \copyright
(2004) IOP Publishing Ltd,
\href{http://www.iop.org/EJ/journal/CQG}{http://www.iop.org/EJ/journal/CQG}}
\author{E. Minguzzi}
 \affiliation{Departamento de Matem\'aticas,
Universidad de Salamanca,  Plaza de la Merced 1, 37008 -
Salamanca, Spain \\ and INFN, Piazza dei Caprettari 70, I-00186
Roma, Italy
\\ minguzzi@usal.es }

\begin{abstract}
Local simultaneity conventions are mathematically represented by
connections on the bundle of timelike curves that defines the
frame. Those simultaneity conventions having an holonomy
proportional to the Riemann tensor have a special interest since
they exhibit a good global behavior in the weak field limit. By
requiring that the simultaneity convention depends on the the
acceleration, vorticity vector and the angle between them we are
able to restrict considerably the allowed simultaneity
conventions. In particular, we focus our study on the simplest
among the allowed conventions. It should be preferred over
Einstein's in flat spacetime and in general if the tidal force
between neighboring particles is weaker than the centrifugal
force. It reduces to the Einstein convention if the vectorial
product between the acceleration and the vorticity vector
vanishes. We finally show how to use this convention in practice.
\end{abstract}

\pacs{04.20.-q, 04.50.+h}
\maketitle


\section{Introduction}
In 1904 Poincar\'e \cite{galgani96,poincare04a,poincare04b}
defined a synchronization convention for distant clocks that is
nowadays know as Einstein synchronization (simultaneity)
convention \cite{galison03}. Given two space points $A$ and $B$
Poincar\'e defined that the clocks at $A$ and $B$ are
``synchronized" if the time it takes light to go from $A$ to $B$
is the same of that needed by light to go from $B$ to $A$. It was
later shown \cite{weylG,minguzzi02d} that this definition is also
transitive between space points if the speed of light over closed
paths is a universal constant $c$. In this case it is possible to
construct a global time variable $t$ and define as simultaneous
those events that correspond to the same value of $t$. These are
actually some steps needed to develop special relativity on a firm
ground. With the introduction of Minkowski spacetime the previous
definition of simultaneity acquired a geometrical interpretation:
two events of coordinates $a^{\mu}$ and $b^{\mu}$ are simultaneous
with respect to the observer (inertial reference frame) of
4-velocity $u^{\mu}$ if and only if $u \cdot (a-b)=0$, that is,
the simultaneity slices belong to the ker of $u_{\mu}\dd x^{\mu}$.
This was a signal that something new was happening in the study
and application of the simultaneity concept.

Actually, in the second half of the nineteenth century there was a
lot of confusion about the  diffusion of time. Many cities had
their own local time and, as reported by Kern \cite{kern83}, in
France there was even a delay of five minutes between the time on
the trains and the time in the railway stations, so as to give to
the passengers five more minutes to get on the trains. After the
introduction of {\em Greenwich Mean Time} (GMT)
\cite{howse80,blaise00}, the synchronization of clocks was related
to GMT  through the definition of time zones. However, as any
general relativist knows, that procedure is meaningless without
giving a well defined notion of GMT in points different from
Greenwich. One can try to use the Einstein convention in order to
achieve a worldwide synchronization but, unfortunately, Sagnac
discovered in 1913 an effect that prevents the Einstein
synchronization convention from being transitive on rotating
frames like the earth \cite{sagnac13,ashtekar75,anandan81}. In
fact, a worldwide syncrhonization would imply a ``time ambiguity''
of the order of one hundred of nanoseconds. It means that clocks
can be synchronized using the Einstein convention but statements
regarding the comparison of distant clocks for time intervals
below or of the order of one hundred of nanoseconds are
meaningless. Indeed at that scale time intervals strongly depend
on the actual succession of  clock synchronizations that have been
followed to coordinate clocks. So, we can say that  event $A$
happened 200 nanoseconds after $B$ only if  events $A$ and $B$
happen at the same point on the earth. The synchronization
convention is not transitive; therefore, there is no exact
definition of Einstein global time. However, one can still use
Einstein synchronization being aware of the limits of the
procedure.

As the clocks became more accurate there was the need of a much
more accurate definition of coordinate time. The Greenwich Mean
Time was replaced by what is now known as {\em Coordinated
Universal Time} (UTC)  and both special relativistic effects, like
the one by Sagnac, and general relativistic effect, like time
dilation due to the gravitational field, were taken into account.
In particular the clock synchronization is achieved using the  GPS
(global positioning system) using a procedure known as ``common
view''(complemented by other less accurate methods, see tables C1
and C2 of \cite{allan97} or \cite{itu97}). It should be said that
the synchronization using this system starts from a model of
spacetime that ignores the general relativistic effect of dragging
of inertial frames \cite{allan84,ashby03,ciufolini95}. That is, it
is assumed the existence of a timelike Killing vector field having
zero vorticity. This Killing vector field defines the inertial
frame at rest with respect to the fixed stars, the so called
Earth-Centered Inertial (ECI) frame, and since there is no
vorticity, the Einstein convention becomes transitive in that
frame. The GPS aims at reproducing a coordinate time  for the
rotating earth so that the simultaneity coincides with the ECI
frame simultaneity \cite{ashby02,ashby03}. We know, however, that
there is no Killing vector field having the assumed properties
since, as the Kerr metric tells us, the Killing field that remains
timelike at infinity has a non-vanishing vorticity
\cite{hawking73}. Of course, this is a small effect that turns out
to be negligible with respect to the actual order of accuracy
required by practical applications, but it shows that, also in the
synchronization through the GPS there is an intrinsic time
ambiguity (using the Kerr metric it turns out to be of the order
of $10^{-16}$s \cite{tartaglia98}) since there is no such exact
thing as Einstein time in the inertial frame. As a consequence, as
we do in the present paper, it is convenient to define the
coordinate time as a function having an intrinsic indeterminacy,
its value being based on a synchronization procedure which is path
dependent. The coordinate time is then defined through the
corresponding synchronization procedure that in general would be
only approximately transitive (small time ambiguity, i.e small
holonomy of the related connection, see \ref{sec1}).

We use the following  notations: The spacetime metric has
signature (+ - - -) and the wedge product is fixed by $\alpha
\wedge \beta=\alpha \otimes \beta - \beta \otimes \alpha$. The
Levi-Civita tensor satisfies, in Minkowski spacetime,
$\varepsilon_{0123}=1$. Sometimes, if no ambiguity arises, we will
denote vectors (or 1-forms) like 4-velocity $u^{\alpha}$, with the
short notation $u$.

One can try to extend the simultaneity concept to general
relativity. First of all, note that Poincar\'e defined the
simultaneity starting from a well defined notion of space. Indeed,
in general, simultaneity is a concept that should be related with
respect to a frame. In general relativity this is done introducing
a congruence of timelike curves generated by a unitary timelike
vector field $u(x)$ (threading). A particle is then defined to be
``at rest" in the frame if its worldline belongs to the congruence
of timelike curves\cite{hawking73,minguzzi03}. The space $S$ is
the set of timelike curves generated by $u(x)$ and the, time
dependent, space metric is
\begin{equation} \label{space}
\dd l^2= -h_{i j} \dd x^{i} \dd x^{j} \qquad i,j=1,2,3
\end{equation}
where $h^{\mu}_{\nu}=\delta^{\mu}_{\nu}-u^{\mu}u_{\nu}$ is a
projector.  The issue of synchronizing  clocks at rest and that of
defining simultaneity between events become distinct in general
relativity. Given a foliation of spacetime, one could define the
events as simultaneous when they lie on the same slice, however,
in general, it is impossible to parametrize the slices in such a
way that the parameter, say $t$, measures up to a constant the
proper time of any clocks at rest in the frame. This is possible
if each hypersurface of the foliation is obtained from a single
hypersurface under a diffeomorphism $\phi_{\tau}$ generated by
$u=\p_{\tau}$. Thus we shall say that clocks at rest are {\em
synchronizable} if after a resetting their proper times coincide
on the simultaneity slices\footnote{Sachs and Wu \cite{sachs77}
define a reference frame to be {\em synchronizable} if the
Einstein simultaneity convention gives rise to a spacetime
foliation. We shall not use the word {\em synchronization} as
Sachs and Wu do. For us clocks on the frame are synchronized if
their proper time coincide on the spacelike slices of the global
simultaneity convention considered. Thus there is no reference to
the Einstein simultaneity convention.}. In general, given the
global simultaneity convention, clocks are not synchronizable even
for simple metrics. The interesting issue becomes, therefore, that
of finding an operational definition of coordinate time (global
simultaneity convention). Consider the newtonian approximation of
earth's metric. In the rotating frame it takes the form
\cite{ashby03}
\begin{equation}
ds^{2}=(1+2(\Phi-\Phi_{0})) \dd t^{2}-(1-2V)[\dd r^{2}+r^{2} (\dd
\theta^{2}+\sin^{2}\theta \dd
\phi^{2})]-2\omega_{E}r^{2}\sin^{2}\theta \, \dd \phi \dd t.
\end{equation}
where $V$ is the gravitational potential, $\Phi$ is the
gravitational plus the centrifugal potential, and $\Phi_{0}$ is
the value of $\Phi$ at the surface of the earth geoid. It is
introduced so that the variable $t$ represents the proper time of
clocks at rest in the earth surface. The frame here is the one
generated by $\p_{t}$ and the question is how to assign in
practice  the label $t$ of the previous formula to the events. As
we said, the Einstein simultaneity convention does not work
accurately because of the Sagnac effect. The GPS was conceived to
do this, but it has some disadvantages that we wish to point out.
Unlike the Einstein convention, the GPS is a non local method of
obtaining a global simultaneity convention. In fact, it depends on
global information like the shape of the geoid, its potential and
the position of the satellites with respect to the earth. One has
to take into account all of these aspects and their influence on
the rate of clocks in the satellites. The Einstein convention is
conceptually easier since it treats on the same footing all the
points at rest in the frame and does not require any global
information in order to be applied. It does not require the
spacetime to have particular symmetries, or the spacetime metric
to be a solution of Einstein equations. Unfortunately, it does not
work accurately in the large but it gives us the hint that these
kind of local simultaneity conventions are worth studying. By {\em
local simultaneity convention} we mean a definition of
simultaneity that applies locally, like a rule between neighboring
observers. Good local simultaneity conventions are those  that are
``almost integrable" in the sense that they give rise globally to
a distribution of horizontal planes which, in the region under
consideration and if the accuracy of experiments is taken into
account, has a negligible holonomy. We thus come to the problem
addressed in this paper: finding a local simultaneity convention
that exhibits a good global behavior. We shall not solve this
problem completely but we will make a first significant step in
that direction.

We mention that some authors \cite{alba03} have argued that the
classical 1+3 (threading) approach, taken up here, runs into
problems because the Einstein convention, in general, is not
transitive. It is for this reason that they propose an alternative
3+1 approach typical of Hamiltonian field theories. The problem,
however, is not with the threading approach but rather with the
Einstein convention. It is possible to consider local conventions
different from Einstein's as we do in the present paper. Moreover,
the threading approach is always in direct contact to observables,
a feature that makes it really appealing. This line of research
has not yet been explored; indeed, the very definition of local
simultaneity convention is given here for the first time.
Ultimately, this operational approach to coordinate time may
provide new insights into the problem of building coordinates in
curved spacetimes for both positioning an navigational purposes
\cite{guinot97,bahder01,rovelli02,blagojevic02}, particularly in
the general case of arbitrary spacetimes.

A word should also be said on the role of constant (or zero) mean
curvature spacelike hypersurfaces. Their study arose naturally in
numerical relativity \cite{eardley79} since the Cauchy problem
simplifies starting from initial data on this kind of
hypersurfaces. Moreover, their evolution was also shown to have
very useful properties like that of avoiding a class of crushing
singularities\cite{eardley79}. Also a number of quite strong
results on the existence of such hypersurfaces and foliations
appeared in the literature \cite{gerhardt83,bartnik88,isenberg98}
so that it would be natural to consider them as a candidate for an
operative definition of simultaneity. However, these hypersurfaces
are not derived from a congruence of timelike curves and therefore
it is not clear how they relate to the frame concept and whether
this relation could be local. For instance, one could try to
define the congruence of timelike curves from a foliation of
constant mean curvature hypersurfaces instead of deriving from a
congruence of timelike curves a spacelike foliation. This approach
could have some advantage but faces with the problem that the
opposite is needed in practical applications and, in general, only
a restricted set of congruences of timelike curves would be
generated in that way. In fact, suppose a constant mean curvature
foliation is given and define the corresponding congruence of
timelike curves as the one generated by the normals to the slices,
then only vortex-free congruences would be generated. This gives
no hint on how to assign a foliation to congruences with
non-vanishing vorticity which is exactly the problem that already
the Einstein convention left unsolved. It is for these reasons
that constant mean curvature slices will not play a significative
role in our study.

Finally, we stress that the determination of a useful simultaneity
concept, i.e. of an operative definition of time function, has no
consequence at the level of physics laws. This follows from the
principle of general covariance which assures that the choice of a
suitable time function can, especially in spacetimes with
symmetries, simplify the physics equations without however
altering their ultimate physical content. Although we shall
advocate that a certain local simultaneity convention should be
preferred, we shall mean by this only that such convention is more
useful in order  to assign a ``time label" to events while at the
level of physics laws  nothing  changes.

\section{Mathematical framework} \label{sec1}
In this work we adopt a fully covariant point of view. We will
make use of the hydrodynamic formalism developed in
\cite{ehlers61} (see also \cite{ellis73}) and of some results of
gauge theories on fiber bundles \cite{kobayashi63,michor91}.

 Let $\pi: M \to S$ be the
differentiable projection that to any event $x$ associates the
timelike curve generated by $u$, $s=\pi(x)$, passing through it.
In this way $M$ acquires a local structure of fibre bundle over
$S$. Locally the simultaneity is represented by a projector field,
\begin{equation}
C(x): TM_{x} \to VM_{x}
\end{equation}
on the vertical space at $x$ (i.e. the one-dimensional space
spanned by $u$). The ker of $C(x)$ selects a horizontal space
$HV_{x}$, $TM_{x}=VM_{x}\oplus HM_{x}$.  By definition, we shall
say that, events on a small neighborhood of $x$ that lie on the
exponential map of $HM_{x}$ are $C$-simultaneous with respect to
the observer of 4-velocity $u$. It can be shown \cite{minguzzi03}
that $C$ is a generalized  connection in the formalism of
generalized gauge theories \cite{mangiarotti84,modugno91}. For
more on the gauge interpretation of simultaneity see
\cite{minguzzi03}.\\

\noindent
 {\bf Def. Local simultaneity convention}. A local simultaneity
 convention is a projector field $C(x): TM_{x} \to VM_{x}$ that
 depends solely on tensors built from $u^{\alpha}$, $g_{\alpha \beta}$,
 $\varepsilon_{\alpha \beta \gamma \delta}$, and whose horizontal planes are spacelike.\\

This definition follows from the fact that a convention must
depend on measurable quantities, and the dependence must be the
same everywhere. In this way, an observer investigating the motion
of neighboring observers and other local spacetime properties can
find all the information required in order to apply the
simultaneity convention. In general the simultaneity connection
$C$ will therefore be built using the 4-velocity vector, the shear
tensor, the expansion scalar, the vorticity tensor, the metric,
the Riemann tensor and the Levi-Civita antisymmetric tensor. More
exotic tensors can,  in principle, enter the construction but they
should have a clear meaning from an operational point of view.

In this framework the Einstein simultaneity convention is
$C^{\mu}_{\ \nu}=P^{\mu}_{\ \nu}\equiv u^{\mu}u_{\nu}$. Its
operational meaning has been discussed in the introduction.

It is convenient to write $C= \omega \otimes u$ , with $\omega$
1-form over $M$, $\omega(u)=1$ and $\omega_{\alpha}
\omega^{\alpha}>0$. Then $v \in HM_{x} \Leftrightarrow
\omega(v)=0$. The curvature $\hat{R}$ of $C$, is a vertical valued
two-form\footnote{The curvature of the simultaneity connection has
an hat in order to be distinguished from the Riemann tensor.}
\cite{michor91},
\begin{equation}
2\hat{R}=-[C, C]_{FN},
\end{equation}
where $[,]_{FN}$ is the Fr\"olicher-Nijenhuis bracket.  After some
calculations, one has more simply
\begin{equation}
\hat{R}= \Omega  \otimes u  \quad \textrm{with} \quad \Omega=(\dd
\omega) C^{\perp} .
\end{equation}
That is, $\hat{R}^{\delta}_{\alpha \beta}=u^{\delta}(\omega_{\nu;
\mu}-\omega_{\mu;
\nu})(\delta^{\mu}_{\alpha}-\omega_{\alpha}u^{\mu})(\delta^{\nu}_{\beta}-\omega_{\beta}u^{\nu})$.
If $\hat{R}=0$ the distribution of horizontal planes is integrable
and the local simultaneity convention is globally well behaved.
Let us introduce
\begin{equation}
v^{\eta}=-\frac{1}{4}h^{\eta}_{\nu} \varepsilon^{\nu \beta \alpha
\gamma} \omega_{\beta} \Omega_{\alpha
\gamma}=\frac{1}{2}h^{\eta}_{\nu}\varepsilon^{\nu \beta \alpha
\gamma} \omega_{\beta} \omega_{\alpha;\gamma},
\end{equation}
it is a kind of generalized  vorticity vector for the present
non-time orthogonal connection. The inverse formula is
\begin{equation} \Omega_{\alpha
\beta}=-2\varepsilon_{\alpha \beta \gamma \delta}u^{\gamma}
v^{\delta}.
\end{equation}
The conditions $\hat{R}=0$, $\Omega=0$ and $v=0$ all state that
the distribution of horizontal planes is integrable. Note that the
Einstein convention is not globally well behaved since $v$ equals
the vorticity vector. What is worse, the non-integrability does
not arise from constraints imposed by the spacetime geometry. A
kinematical property alone as rotation prevents the simultaneity
convention from working in the large. This fact makes it clear
that a better convention would be welcome. Observe that the
condition $v=0$, that is $h^{\eta}_{\nu} \varepsilon^{\nu \beta
\alpha \gamma} \omega_{\beta} \Omega_{\alpha \gamma}=0$ is
equivalent to the more usual Frobenius condition $\varepsilon^{\nu
\beta \alpha \gamma} \omega_{\beta} \Omega_{\alpha \gamma}=0$,
indeed  $z^{\nu}=\varepsilon^{\nu \beta \alpha \gamma}
\omega_{\beta} \Omega_{\alpha \gamma}$ is orthogonal to
$\omega_{\nu}$ and since $\omega_{\nu}$ is timelike, $z^{\nu}$ is
spacelike and then $z^{\nu}=0$ iff $h^{\eta}_{\nu}z^{\nu}=0$.

The ker of $\omega$ coincides with the ker of
\begin{equation}
u'_{\alpha}=(\omega^{\beta}\omega_{\beta})^{-1/2}\omega_{\alpha} ,
\end{equation}
which can be interpreted as a 4-velocity field ${u'}^{\alpha}
u'_{\alpha}=1$. Hence {\em the C-simultaneity convention is
equivalent to the Einstein convention with respect to observers
moving with a 4-velocity $u'$}.

\subsection{Stationary case}
 If the congruence of timelike curves is
generated by a Killing vector field, another way to introduce a
curvature follows from the mathematical formulation of gauge
theories as connections on  principal fiber bundles. Indeed over
$M$ acts the one parameter group $T_{1}$ of translations
$\phi_{t}$ generated by the Killing vector field $k$, $u=k/\sqrt{k
\cdot k}$. Locally one recovers the structure of a principal fiber
bundle, and the base $S$ can be interpreted as the quotient of $M$
over $T_{1}$. Let $\chi=\sqrt{k \cdot k}$  then, given a local
simultaneity convention, consider the 1-form
\begin{equation} \label{haha}
\sigma= \chi^{-1} \omega  .
\end{equation}
It is clear that (a) $\sigma(k)=\chi^{-1}\omega(k)=1$, and that
(b) $L_{k}\sigma=0$. Indeed, in order to prove (b) note that
$L_{k}\chi=0$ and $L_{k} \omega=0$ since $\omega$ depend on
$u^{\alpha}$, $g_{\alpha \beta}$, $\varepsilon_{\alpha \beta
\gamma \delta}$, and these tensors have a vanishing Lie
derivative. Any 1-form field over the fiber bundle $M$ that
satisfies (a) and (b) is by definition a connection in the sense
of principal bundle gauge theories \cite{kobayashi63}. It is only
a slightly different way of defining a connection with respect to
that of generalized gauge theories
\cite{mangiarotti84,modugno91,michor91}. Whatever definition is
used, the connection always defines a distribution of horizontal
planes given by $\textrm{ker}\,\sigma=\textrm{ker}\,\omega$. The
curvature is defined by
\begin{equation}
\Sigma_{\mu \nu}=(\dd \sigma)_{\mu \nu} .
\end{equation}
We use the formula $L_{k}=\dd i_{k}+i_{k} \dd$ on $\sigma$
\begin{equation}
L_{k}\sigma=\dd \sigma(k)+i_{k} \dd \sigma ,
\end{equation}
but the first two terms vanish and therefore $\Sigma_{\mu
\nu}u^{\nu}=0$. We search a relation between the two definitions
of curvature $\Sigma$ and $\Omega$. Taking the exterior derivative
of $\chi^{-1}\omega$ and projecting on the horizontal plane we
obtain
\begin{equation}
(\dd \sigma)C^{\perp}=(\dd \chi^{-1})C^{\perp} \wedge \omega
C^{\perp}+\chi^{-1}\dd \omega C^{\perp} ,
\end{equation}
but $\omega C^{\perp}=0$ and since $i_{u}\dd \sigma=0$, we have
$(\dd \sigma) C^{\perp}=\dd \sigma$, and finally,
\begin{equation}
\Omega=\chi \Sigma.
\end{equation}
$\Sigma=\dd \sigma=0$ is another integrability condition for the
distribution of horizontal planes. It is particularly convenient
in this form since it implies that (this is always true locally
and holds globally in a simply connected spacetime) there is a
time function $t: M \to \mathbb{R}$ such that $\sigma=\dd t$. We
shall call this time, {\em $C$-Killing time}, since on a worldline
of the reference frame the rate of this time equals the one given
by the Killing vector field ($\dd t(k)=1$) and the hypersurfaces
of simultaneity are those of the C-simultaneity
($\textrm{ker}\,\dd t=\textrm{ker}\,\omega$).

\section{Stationary frames}
There is little hope of finding a globally well behaved
simultaneity convention that works in every case. In the present
paper we tackle the problem introducing some restrictions
\begin{itemize}
\item Stationarity. We assume this property mainly for technical
reasons: it allows us to use repeatedly the Killing vector lemma
and to simplify the expressions. Thus, we shall assume that the
congruence of timelike curves is generated by a timelike Killing
vector field $k$. This assumption is not, however, so restrictive.
Firstly, we are mostly interested on simultaneity conventions to
be applied at the exterior of massive objects. Most of them are
already studied in the stationary case i.e. when the dynamical
effects have come to an equilibrium.  Secondly, the concept of
frame is particularly sharp and clear in the stationary case since
there, even a time independent space metric can be defined. It is,
therefore, a natural starting point for more complicated
investigations. \\
\item $v^{\alpha} \ne 0$. We relax the
integrability condition $v^{\alpha}=0$ since, in practice, we need
conventions that lead to distributions of horizontal planes which
are approximately integrable at least over the scales where a
global time variable is required. To this end we weaken that
condition looking for conventions having a curvature proportional
to the Riemann tensor
\begin{equation}
v^{\alpha} \propto R^{\gamma}_{\ \delta \eta \mu \nu}.
\end{equation}
In this way at least in a weak field limit the distribution of
horizontal planes is integrable.  The holonomy of a connection is
proportional to the area enclosed by a closed curve and the
curvature. Over a length scale $L$ it is given by
\begin{equation}
\Delta \sim  L^{2}\mathcal{R}T ,
\end{equation}
where $\mathcal{R}$ is the typical value of the non-zero Riemann
coefficients and $T$ is the typical value for the tensor which
contracts with the Riemann tensor to give the curvature $v$. This
should be compared with the relation that follows from the
Einstein convention
\begin{equation}
\Delta \sim  L^{2} w .
\end{equation}
Thus we are looking for a connection that, at least in the weak
field limit $\mathcal{R} \to 0$, is much more well behaved than
Einstein's. Moreover, the lack of integrability here is caused by
the spacetime geometry (Riemann tensor) and not by the kinematical
properties of the reference frame (vorticity). In this sense the
lack of integrability is much more natural and acceptable.
\end{itemize}

We recall some basic definitions that will be needed in the
statement of a theorem. Let $k$ be the timelike Killing vector
field. The Killing vector lemma\cite{geroch71} states that the
following relation holds
\begin{equation}
k_{\alpha ; \beta ; \gamma}= - R_{\gamma \delta \alpha \beta}
k^{\delta} .
\end{equation}
For a timelike Killing vector field the 4-velocity is $
u^{\alpha}={k^{\alpha}}/{\sqrt{k \cdot k}}$, the acceleration is
\begin{equation}
a^{\alpha}=u^{\alpha}_{; \beta} u^{\beta}= -(\ln\chi)_{; \beta}
h^{\beta \alpha}=-(\ln\chi)^{; \alpha} ,
\end{equation}
and the shear tensor and the expansion scalar vanish. Let us
define (this is Eq. (\ref{haha}) for the Einstein convention),
$\sigma^{e}=k/(k\cdot k)$, the vorticity tensor can be written
\begin{eqnarray*}
w_{\alpha \beta} &=& u_{[\gamma ;\delta]} h^{\gamma}_{\alpha}
h^{\delta}_{\beta}=\frac{k_{\alpha ; \beta}}{\sqrt{k \cdot
k}}+[u_{\alpha} a_{ \beta}-u_{\beta} a_{ \alpha} ]= \chi
\sigma^{e}_{[\alpha ; \beta]} ,
\end{eqnarray*}
and the vorticity is defined as $w^{\alpha}=\frac{1}{2}
\epsilon^{\alpha \beta \gamma \delta} u_{\beta} w_{\gamma \delta}$
 having inverse formula $w_{\alpha \beta}=\epsilon_{ \alpha \beta \gamma \delta}
u^{\gamma} w^{\delta}$. Finally, we define the 1-form field
$m_{\alpha}=w_{\alpha \beta} a^{\beta}$. With the short notation
$a^{2}$ we denote the positive number $a^{2}=-a^{\alpha}
a_{\alpha}$, and analogously for $w^{\alpha}$ and $m_{\alpha}$.
The vector $m^{\alpha}$ is perpendicular to $u^{\alpha}$,
$w^{\alpha}$ and $a^{\alpha}$. The tetrad
$\{u^{\alpha},w^{\alpha},a^{\alpha},m^{\alpha}\}$ is a base in
those events where $m^{\alpha} \ne 0$. Note  that $m^{\alpha} \ne
0 \Leftrightarrow m^{2} \ne 0$ where $m^{2}=a^{2} w^{2} \sin^{2}
\theta$ and $\theta$ is the angle between the acceleration and the
vorticity vector. In the same events any local simultaneity
convention takes the form
\begin{equation}\label{generic}
\omega_{\alpha}=u_{\alpha}+\psi^{m}(x) m_{\alpha}+\psi^{a}(x)
a_{\alpha}+ \psi^{w}(x) w_{\alpha} ,
\end{equation}
for suitable functions $\psi^{m},\psi^{a},\psi^{w}$. From the
definition of local simultaneity convention it follows moreover
that $\psi^{m},\psi^{a},\psi^{w}$, depend on the acceleration $a$
the vorticity $w$, the angle $\theta$, and on more exotic scalars
(note that in a stationary frame the possibilities are reduced
since the shear and the expansion vanish). The following theorem
explores the consequences of a dependence on $a$, $w$ and
$\theta$. \\

 {\bf Theorem.} In a stationary spacetime let $k$ be a timelike
Killing vector field and set $u=k/\sqrt{k \cdot k}$. Let $U$ be
the open set $U=\{x: m(x)
> 0 \ \textrm{and} \ a(x) \ne w(x) \}$. Consider in $U$ the 1-form
\begin{equation} \label{conn}
\omega_{\alpha}=u_{\alpha}+\psi^{m}(x) m_{\alpha}+\psi^{a}(x)
a_{\alpha}+ \psi^{w}(x) w_{\alpha}.
\end{equation}
\begin{itemize}
\item[(i)] Let $\psi^{m},\psi^{a},\psi^{w}$, be $C^{1}$ functions dependent only
on $a$, $w$ and $\theta$.  Then, regardless of the stationary
spacetime considered, the connection is timelike in $U$ (and hence
it is a simultaneity connection in $U$) and has a curvature
proportional to the Riemann tensor in $U$  if and only if
\begin{equation} \label{magic}
\psi^{m}=\bar{\psi}^{m} \equiv \frac{a^{2}+w^{2}-\sqrt{(a^{2}+w^{2})^{2}-4m^{2}}}{2m^{2}}, \\
\end{equation}
and there is a $C^{1}$ function $b: \mathbb{R} \to \mathbb{R}$ of
the variable $(a w \cos \theta)/(a^{2}-w^{2})$ such that for $m>0$
and $a \ne w$ the following inequality holds
\begin{equation} \label{inequal}
\frac{\sqrt{(a^{2}+w^{2})^{2}-4m^{2}}-(a^{2}-w^{2})}{2w^{2}} \,
b^{2}+\frac{m^{2}}{w^{2}} \psi^{a 2}+m^{2} \bar{\psi}^{m 2} <1 ,
\end{equation}
and
\begin{equation} \label{w}
\psi^{w}=-\frac{a \cos \theta}{w} \,\psi^{a}+\frac{b}{w^{2}} \{
\frac{\sqrt{(a^{2}+w^{2})^{2}-4m^{2}}-(a^{2}-w^{2})}{2}\}^{1/2}.
\end{equation}
Thus, in order to have a simultaneity connection on $U$ having a
curvature proportional to the Riemann tensor, it suffices to
choose a pair $(\psi^{a},b)$ so as to satisfy the inequality
(\ref{inequal}). Indeed $\psi^{m}$ is fixed by (\ref{magic}) while
$\psi^{w}$ follows from $b$ and $\psi^{a}$.
 \item[(ii)] Consider in $U$  the local simultaneity
convention defined by $\psi^{a}=\psi^{w}=0$
\begin{equation} \label{cbar}
\bar{\omega}_{\alpha}=u_{\alpha}+
\frac{a^{2}+w^{2}-\sqrt{(a^{2}+w^{2})^{2}-4m^{2}}}{2m^{2}}\,
m_{\alpha},
\end{equation}
It is timelike, i.e. satisfies (\ref{inequal}), and can be
extended by continuity to the set $C=A-B$ where, $A=\{x:
a^{2}+w^{2}>0\}$, $B=\{x: a=w\ne 0 \ \textrm{and} \
\theta=\pi/2\}$, by defining, $\bar{\omega}_{\alpha}=u_{\alpha}$,
in those points where $m=0$. Its curvature is
\begin{eqnarray}\label{dfg}
\bar{v}^{\eta}&=&\frac{\bar{\psi}^{m}}{2}h^{\eta}_{\ \nu}
\varepsilon^{\nu \beta \alpha \gamma} u_{\beta} R_{\gamma \delta
\mu \sigma} \{
\frac{m_{\alpha}}{\sqrt{(a^{2}+w^{2})^{2}-4m^{2}}}(u^{\delta}
w^{\mu \sigma}+2u^{\delta} u^{\mu} a^{\sigma}) \nonumber \\  && \
+m_{\alpha}
\frac{2\bar{\psi}^{m}}{\sqrt{(a^{2}+w^{2})^{2}-4m^{2}}}(u^{\delta}m^{\mu}a^{\sigma}+u^{\mu}u^{\delta}w^{\sigma}_{\
 \tau}m^{\tau})+u^{\mu}u^{\delta}w_{\alpha}^{\
\sigma}-\delta^{\mu}_{\alpha}u^{\delta}a^{\sigma} \} .
\end{eqnarray} \\
\end{itemize}
Motivated by (ii)  we shall focus our study on the following
simultaneity convention \\

{\bf Def.} The $\bar{C}$-simultaneity convention is the convention
(\ref{cbar}) defined on the set $C=A-B$. \\

Note that the conditions imposed on the sets in the statement of
the theorem are not restrictive and arise only for technical
reasons. For instance, the set $B$ is empty in most cases and does
not have any particular physical interpretation. In practical
applications, the $\bar{C}$-simultaneity will be defined on the
whole set $A=\{x: a^{2}+w^{2}>0\}$. Moreover, in most cases, the
set $A$ will be the entire spacetime. Indeed, for spacetimes
having $w=0$ on open sets the search for an integrable
simultaneity convention does not make much sense since the
Einstein simultaneity convention is already integrable. In the
theorem there are therefore only four physical assumptions: the
spacetime is stationary; the connection scalars depend only on
$a$, $w$ and $\theta$ (it means that in order to apply the
convention only the acceleration and the vorticity should be
locally measured); the connection  has spacelike horizontal planes
(otherwise it would not be a simultaneity connection); its
curvature is proportional to the Riemann tensor (it allows to
ignore the holonomy in the weak field limit).

Remarkably, $\bar{C}$-simultaneity follows almost uniquely from
these geometrical requirements. While the observers have the
freedom to choose a local simultaneity convention, the requirement
of being almost integrable in the weak field limit imposes strong
constraints on its actual expression. The geometry tells the
observers what simultaneity convention is better to use in
practice and remarkably the simultaneity convention that turns out
is not Einstein's but rather $\bar{C}$-simultaneity with possibly
non-vanishing acceleration and vorticity terms.

By construction the curvature is proportional to the Riemann
tensor. In order to understand the physical implications of Eq.
(\ref{dfg}) some comments are in order. It is not difficult to see
that the following inequalities hold in $C$
\begin{eqnarray}
m \bar{\psi}^{m}&<&1 , \\
(a^{2}+w^{2})\bar{\psi}^{m}&<&2.
\end{eqnarray}
Using them  and defining $v^{2}=-v^{\eta} v_{\eta}$, from the
expression for $\bar{v}^{\eta}$ we find that
\begin{equation}
\bar{v} \lesssim 10
\mathcal{R}\frac{\sqrt{a^{2}+w^{2}}}{|a^{2}-w^{2}|} ,
\end{equation}
where $\mathcal{R}$ is the value of the greatest non vanishing
component of the Riemann tensor in a orthonormal base. Restoring
$c$ one sees that $a^{2}$ is suppressed by an additional factor
$c^{2}$ with respect to $w^{2}$. In  typical practical situations
having a non-vanishing vorticity, it therefore happens that $w^{2}
\gg a^{2}$ and the previous equation reads
\begin{equation}
\bar{v} \lesssim 10 \frac{\mathcal{R}}{w}
\end{equation}

We recall that $v=w$ for the Einstein convention, thus in this
case the $\bar{C}$-simultaneity connection is better than
Einstein's if
\begin{equation}
\mathcal{R}  \ll w^{2} .
\end{equation}
Now, the force experienced between two points at rest in the frame
is expected to have a tidal component of the form $\mathcal{R}
\delta x$, where $\delta x$ is the displacement, and a centrifugal
component of the form $w^{2} \delta x$. Thus, roughly speaking,
{\em the $\bar{C}$-simultaneity convention is better than
Einstein's in those spacetime regions where tidal forces are
weaker than centrifugal forces}.

We immediately see that it is not convenient to use the
$\bar{C}$-simultaneity at the surface of the earth or at the
surface of a spinning planet. In fact $\mathcal{R}$ is, in
geometrized units, of the order of $M/R^{3}$, where $R$ is the
planet radius. Consider a body on the surface, at a distance $d
\sim R \sin \theta$ from the axis of rotation. The component of
the gravitational force directed towards the axis is greater than
the centrifugal force. The classical equilibrium of forces gives
\begin{equation}
\frac{M}{R^{2}}\sin\theta > w^{2}d ,
\end{equation}
otherwise the body would not stay on the surface of the planet.
This last equation implies $w^{2} < \mathcal{R}$ and  therefore
the Einstein convention should still be preferred on the surface
of a spinning planet. The $\bar{C}$-simultaneity appears to be a
very good convention but it should be applied only in regions
where the gravitational field is particularly weak. It is
convenient to use the $\bar{C}$-simultaneity at a radius greater
than  that of geostationary orbits. In that region, as the radius
increases, the holonomy over a path $\gamma$ of constant radius
around the planet decreases. Indeed, the holonomy  scales as
$(M/(wR^{3})) S $, where $S$ is the area of a surface such that
$\p S=\gamma$, but $S \sim 2 \pi R^{2}$ and hence the holonomy
scales as $M/(wR)$. On the contrary, the holonomy of the Einstein
convention increases since it scales as $w R^{2}$.

We have calculated the curvature with the assumption of
stationarity but any local simultaneity convention makes sense
even without that assumption. Thus we can consider
$\bar{C}$-simultaneity even in non-stationary cases. It remains
well defined and timelike as long as $w,a,$ and  $\theta$ define a
point in the set $C$. In flat spacetime it should be preferred
over Einstein's. In fact, if the curvature of the Einstein
convention vanishes in a open set $O$, $w=0$, the same happens for
the convention  $\bar{C}$, since in $O$ the two conventions
coincide. On the contrary, there are cases like the rotating
platform such that the curvature of $\bar{C}$  vanishes but $w
\ne0$.

\section{A simple example: the rotating platform}
The theorem implies that in flat spacetime the horizontal planes
of  $\bar{C}$ are exactly integrable. As a consequence stationary
frames in flat spacetime bring an intrinsic privileged global
simultaneity convention. This  feature is quite surprising. In the
gauge interpretation of simultaneity  any global simultaneity
convention is a section of the principal bundle and therefore we
expect global simultaneity conventions to be equivalent since we
can change from one to the other by gauge
transformations\cite{minguzzi03}. However, it should be taken into
account that here we have a special ingredient given by the scalar
field $\chi$. Using this scalar field one is able to construct an
acceleration field and finally the $\bar{C}$-simultaneity. In
general a field like $\chi$ appears in those theories which have
the mathematical structure of Kaluza-Klein theories with a
non-constant scalar field, actually the case
here\cite{minguzzi03}.

It is natural to check the property of $\bar{C}$-simultaneity in
the rotating platform case. The rotating platform has always
provided a natural theoretical laboratory for the study of
rotating spacetimes.  There is, for instance, a wide array of
literature on the definition of space metric \cite{gron03} and on
the problem of clock synchronization \cite{anderson98,minguzzi02}
for the rotating platform metric. The rotating platform is also
interesting in practice since the Sagnac effect \cite{stedman97}
can be recognized as a universal effect of rotating frames
starting from an intrinsic description of the effect in the
rotating platform case. It is nothing but the manifestation of the
non-vanishing holonomy for the Einstein simultaneity
convention\cite{minguzzi03}. Our task here is to show that
observers on a rotating platform using the $\bar{C}$-simultaneity
convention succeed in constructing a global coordinate time. As
will be explained in detail in section \ref{practice}, the
operational procedure associated to a local simultaneity
connection, as the name suggests, is local and therefore the
observers on the rotating platform may even ignore their being on
a rotating platform. This does not prevent the application of
$\bar{C}$-simultaneity convention from succeeding.

The metric is obtained from the flat metric in cylindric
coordinates,
\begin{equation}
\dd s^{2}=\dd t^{2}-\dd \rho^{2}-\rho^{2} \dd \theta^{2} -\dd
z^{2} ,
\end{equation}
 after the change of coordinates, $\theta \to \theta + \omega t$, where $c=1$ and $\omega$ is a constant
\begin{equation}
\dd s^{2}=(1-\rho^{2}\omega^{2})\dd t^{2}-\dd \rho^{2}-\rho^{2}
\dd \theta^{2}-2\rho^{2}\omega \dd \theta \dd t -\dd z^{2},
\end{equation}
or
\begin{equation}
\dd s^{2}=(1-\rho^{2}\omega^{2})(\dd t-\frac{\rho^{2}
\omega}{1-\rho^{2}\omega^{2}} \dd \theta)^{2}-\dd
\rho^{2}-\frac{\rho^{2}\dd \theta^{2}}{1-\rho^{2}\omega^{2}} -\dd
z^{2}.
\end{equation}
 Worldlines of equation $x^{i}=const.$ belong to the congruence
of timelike curves, and $\p_{t}$ is the timelike Killing vector
field. We shall limit ourselves to the open set $\rho \omega <1$,
since in this set the Killing vector field is  timelike.

The scalar field is $\chi^{2}=(1-\rho^{2}\omega^{2})$, and the
acceleration is
\begin{eqnarray}
a_{\alpha}\dd x^{\alpha}&=&-\frac{1}{2}\dd \ln
\chi^{2}=\frac{\omega^{2}\rho}{1-\rho^{2}\omega^{2}} \dd \rho, \\
\end{eqnarray}
its norm being $a^{2}={\omega^{4}
\rho^{2}}/{(1-\rho^{2}\omega^{2})^{2}} $. From the metric we find
\begin{equation}
\sigma^{e}_{\alpha}\dd x^{\alpha}=\dd t-\frac{\rho^{2}
\omega}{1-\rho^{2}\omega^{2}} \dd \theta,
\end{equation}
 the vorticity tensor is
\begin{equation}
w_{\alpha \beta} \dd x^{\alpha} \otimes \dd
x^{\beta}=-\frac{\chi}{2} \dd \sigma=\frac{\omega
\rho}{(1-\rho^{2}\omega^{2})^{3/2}} \dd \rho \wedge \dd \theta,
\end{equation}
and using $g^{\theta \theta}=-(1-\rho^{2} \omega^{2})/\rho^{2}$,
we obtain $ w^{2}={\omega^{2}}/{(1-\rho^{2} \omega^{2})^{2}}$. The
1-form $m_{\alpha}$ is
\begin{eqnarray}
m_{\alpha} \dd x^{\alpha}&=&\frac{\omega^{3}
\rho^{2}}{(1-\rho^{2}\omega^{2})^{5/2}} \dd \theta,\\
.
\end{eqnarray}
its norm being $m^{2}={\omega^{6}\rho^{2}}/{(1-\rho^{2}
\omega^{2})^{4}} $. We see immediately that $m^{2}=a^{2}w^{2}$
hence the angle between $a$ and $w$ is $\pi/2$. In this case the
formula for $\bar{\psi}^{m}$ simplifies to
\begin{equation}
\bar{\psi}^{m}=\frac{\min(a^{2},w^{2})}{m^{2}} ,
\end{equation}
but since in the region under consideration $\rho \omega<1$ we
have $\min(a^{2},w^{2})=a^{2}$.
 Using these formulas we obtain finally,
\begin{equation}
\bar{\omega}_{\alpha}\dd x^{\alpha}=u_{\alpha} \dd
x^{\alpha}+\bar{\psi}^{m}m_{\alpha} \dd x^{\alpha}=\chi \dd t,
\end{equation}
or
\begin{equation}
\bar{\sigma}=\dd t .
\end{equation}
 Thus $\bar{C}$-simultaneity coincides with  Einstein
simultaneity of the inertial frame we started with and the
$\bar{C}$-Killing time is $t$. The hypersurfaces of simultaneity
have equation $t= const.$ From a local simultaneity convention we
have been able to recover the most natural global simultaneity
convention for the rotating platform.

\section{Rotating pseudocylindrical coordinates}

In this section we consider a non trivial example in flat
spacetime. An integral curve of a Killing vector field can be
described, without reference to the neighboring integral curves,
using the Serret-Frenet equation. Because of the translational
invariance of the  spacetime, the acceleration, torsion and
hypertorsion must be constant in absolute value. Curves having
this property have been classified, for the case of flat
spacetime, by Synge \cite{synge67} and Letaw \cite{letaw81}. Then
Letaw and Pfautsch \cite{letaw82} found, for any Killing vector
field in flat spacetime, an adapted coordinate system in the sense
that $k=\p_{t}$. As it is well known, a gauge transformation
transformation $t \to t + \phi(x^{i})$ preserves the relation
$k=\p_{t}$. Letaw and Pfautsch fix the gauge so as to simplify the
expression of the metric. Their choice of coordinate time implies
a simultaneity convention, thus their simultaneity convention
arises from a criteria of simplicity of the metric element. The
less trivial Killing vector field having $m \ne 0$ that they
consider leads to the {\em rotating pseudocylindrical
coordinates}. The
 Minkowski metric in these coordinates is
\begin{equation}
\dd s^{2}=(\xi^{2}-r^{2}\Omega^{2})\dd \tau^{2}-\dd \xi^{2} - \dd
r^{2}- r^{2}\dd \phi^{2}-2r^{2}\Omega \dd \phi \dd \tau ,
\end{equation}
and $k=\p_{\tau}$. It can be rewritten in the form
\begin{equation}
\dd s^{2}=(\xi^{2}-r^{2}\Omega^{2})(\dd \tau
-\frac{r^{2}\Omega}{\xi^{2}-r^{2}\Omega^{2}}\dd \phi)^{2}-\dd
\xi^{2} - \dd
r^{2}-(r^{2}+\frac{r^{4}\Omega^{2}}{\xi^{2}-r^{2}\Omega^{2}}) \dd
\phi^{2} .
\end{equation}
We restrict ourselves to the open set where
$\chi^{2}=\xi^{2}-r^{2}\Omega^{2}>0$, since there the Killing
vector field is timelike.  The acceleration is
\begin{equation}
a_{\alpha} \dd x^{\alpha}=-\frac{1}{2} \dd \ln
\chi^{2}=\frac{\Omega^{2}r}{\xi^{2}-r^{2}\Omega^{2}}\dd
r-\frac{\xi}{\xi^{2}-r^{2}\Omega^{2}} \dd \xi ,
\end{equation}
its norm being
$a^{2}={(\Omega^{4}r^{2}+\xi^{2})}/{(\xi^{2}-r^{2}\Omega^{2})^{2}}$.
From the metric we also easily see that
\begin{equation}
\sigma^{e}_{\alpha} \dd x^{\alpha} =\dd \tau
-\frac{r^{2}\Omega}{\xi^{2}-r^{2}\Omega^{2}}\dd \phi .
\end{equation}
The vorticity tensor is
\begin{equation}
w_{\alpha \beta} \dd x^{\alpha} \otimes \dd
x^{\beta}=-\frac{\chi}{2} \dd \sigma=\frac{\Omega
r\xi^{2}}{(\xi^{2}-r^{2}\Omega^{2})^{3/2}}\dd r \wedge \dd
\phi-\frac{\Omega r^{2}\xi}{(\xi^{2}-r^{2}\Omega^{2})^{3/2}}\dd
\xi \wedge \dd \phi ,
\end{equation}
and taking into account that  $g^{\phi
\phi}=-(\xi^{2}-r^{2}\Omega^{2})/(r^{2}\xi^{2})$, we find
$w^{2}={\Omega^{2}(\xi^{2}+r^{2})}/{(\xi^{2}-r^{2}\Omega^{2})^{2}}
$. The only non-vanishing component of $m_{\alpha}$ is
\begin{equation}
m_{\phi}=\frac{\Omega
r^{2}\xi^{2}}{(\xi^{2}-r^{2}\Omega^{2})^{5/2}}(\Omega^{2}+1) ,
\end{equation}
and its norm is $ m^{2}={\Omega^{2}r^{2}\xi^{2}}
(\Omega^{2}+1)^{2}/{(\xi^{2}-r^{2}\Omega^{2})^{4}}$. Plugging
these results into the expression for $\bar{\psi}^{m}$ we find
$\bar{\psi}^{m}m_{\phi}=\chi^{-1}\Omega r^{2}$ and finally
\begin{equation}
\bar \omega_{\alpha} \dd x^{\alpha}= u_{\alpha} \dd
x^{\alpha}+\bar{\psi}^{m}m_{\alpha} \dd x^{\alpha}=\chi \dd \tau .
\end{equation}
Hence, the $\bar C$-Killing time is the parameter $\tau$ chosen by
Letaw and Pfautsch. This coincidence arises from the fact that the
$\bar C$-simultaneity convention is a natural one. It follows
solely from the metric and the Killing vector field,  so it is
expected to simplify the metric in a coordinate system adapted to
$k$  whose time coordinate induces the $\bar C$-simultaneity
convention. Since Letaw and Pfautsch looked for a gauge that
simplifies the metric it is not surprising that they chose the
$\bar C$ gauge.

\section{Space-Time splitting and gauge fixing}
Usually stationary metrics are written in the form
\begin{equation}
\dd s^{2}=\chi^{2}(\dd t+A_{i}(x^{j})\dd x^{i})^{2}-\dd l^{2} ,
\end{equation}
where $\dd l^{2}$ is the space metric (\ref{space}), $k=\p_{t}$,
and $x^{i}$ are coordinates on $S$. The field $A_{i}$ is not
uniquely determined because of the gauge transformation
\begin{eqnarray}
t &\to & t+\phi(x^{i}) , \label{phi}\\
A_{i} &\to & A_{i}-\p_{i}\phi .
\end{eqnarray}
In a region having a small Riemann tensor we suggest to write the
metric in the following form
\begin{eqnarray}
\dd s^{2} &=& (u_{\alpha}\dd x^{\alpha})^{2}-\dd l^{2} \nonumber\\
&=& \chi^{2}(\bar{\sigma}-\chi^{-1}\bar{\psi}^{m} m_{i} \dd
x^{i})^{2}-\dd l^{2} .
\end{eqnarray}
If the holonomy is sufficiently small we can write, with abuse of
notation, $\bar{\sigma}=\dd t$, where $t$ is the $\bar{C}$-Killing
time. Of course, this equation is understood as an approximation
as long as the Riemann tensor differs from zero. Finally,
\begin{eqnarray} \label{fgh}
\dd s^{2} &=& \chi^{2}(\dd t-\chi^{-1}\bar{\psi}^{m} m_{i} \dd
x^{i})^{2}-\dd l^{2} , \qquad \textrm{(Approximation)} .
\end{eqnarray}
Hence, we have obtained a gauge fixing of the simultaneity
freedom. The gauge potential becomes
\begin{equation}
A_{i}(x^{j})=-\chi^{-1}\bar{\psi}^{m} m_{i} .
\end{equation}
The advantage of Eq. (\ref{fgh}) is that $t$, being the
$\bar{C}$-Killing time, is defined through a local simultaneity
convention and therefore has a clear operational meaning.  In
theory, any global simultaneity convention can be used by the
observers: the change between different global simultaneity
convention  being given by Eq. (\ref{phi}). However, the observers
should also study how to characterize a global simultaneity
convention operationally. We argued that convenient procedures
make use of an almost integrable local simultaneity convention.
This select a global simultaneity convention among the many in
principle available thus leading to the gauge fixing studied in
this section. We shall see in the next section how to find, in
practice, the value of  $C$-Killing time for a given event of
spacetime.

\section{Coordinate time in practice} \label{practice}

In this section we describe how to build a coordinate time in a
stationary spacetime using a $C$-simultaneity connection, the
letter $C$ standing for the convention adopted.  The procedure
will be entirely local, so we have only to describe what an
observer $O$ at rest in the frame has to do, and which information
$O$ needs from neighboring observers in order to assign the
correct coordinate time to the events happening in his/her
worldline. We assume that in the spacetime region under
consideration the Riemann tensor is sufficiently small so that the
holonomy becomes negligible. The coordinate time we wish to build
with the local procedure is therefore the $C$-Killing time. We
also assume $m \ne 0$ in the region under consideration. The
1-form defining the $C$ simultaneity convention can be written
\begin{equation}
\omega_{\alpha}=u_{\alpha}+\psi n_{\alpha}
\end{equation}
where both the scalar field $\psi(x)$ and the normalized spacelike
vector field $n_{\alpha}(x)$ are observables depending on the
acceleration, vorticity, Riemann tensor and on other local
observables.  Note that the Killing vector field $k=\p_{t}$ is
defined only up to a constant factor. As a consequence, the
$C$-Killing time $t$ we wish to recover from local measurement is
defined only up to an affine transformation $t \to at+b$. To
remove this arbitrariness, we fix its value in a point $p$ of
space by saying that in that point $C$-Killing time coincide with
proper time. Of course, the choice of $p$ is arbitrary. Consider a
light beam sent from $p$ to a point $q$ of $S$. The observer in
$q$ detects a redshift($\chi(p)=1$)
\begin{equation}
1+z(p,q)=\chi(q) .
\end{equation}
Thus the $C$-Killing time of an event happening in $q$'s worldline
is related the the proper time of a clock in $q$ by
\begin{equation}
t=\frac{\tau+b(q)}{1+z(p,q)}
\end{equation}
the constant $b$ depending on the synchronization of $q$'s clock.
Let $q' \in S$ be another point in a small neighborhood of $q$.
Let $Z^{\alpha}$ be their spacelike infinitesimal
displacement($Z\cdot u=0$ and $q'$ is the exponential map of $Z$
of base point $q$). Since the frame is stationary, $L_{k}Z=0$. Now
suppose the $C$-simultaneity is almost integrable so that we can
ignore the non-vanishing holonomy if the sensitivity of the
measurements apparatus is taken into account. It means that there
is a natural gauge fixing for this convention analogous the the
one studied for the $\bar{C}$-simultaneity connection in the
previous section
\begin{eqnarray} \label{asdf}
\dd s^{2} &=& \chi^{2}(\dd t-\chi^{-1} \psi n_{i} \dd
x^{i})^{2}-\dd l^{2} , \qquad \textrm{(Approximation)} .
\end{eqnarray}
Here $\dd l$ is the radar distance and its value $D=\sqrt{\dd
l(Z,Z)}$ between $q$ and $q'$ can be easily measured  in the usual
way by sending a light beam in a round trip between the two
points, and by taking half the round trip (proper) time. Now
consider a light beam sent from $q$ to $q'$. From the previous
equation we find that the $C$-Killing interval time the light beam
needs to reach $q'$ is
\begin{equation}
\delta t= \frac{D+\psi n_{i}Z^{i}}{\chi(q)}.
\end{equation}
In terms of proper time ($b=0$)
\begin{equation}
\delta t=\frac{\delta \tau}{\chi}-\frac{\tau}{\chi^{2}} \delta
\chi
\end{equation}
thus
\begin{equation}
\delta \tau=D+\frac{\tau}{\chi} \delta \chi+\psi n_{i}Z^{i}
\end{equation}
Let us introduce the normalized vector $\hat{z}$ such that $Z=D
\hat{z}$, and call $\alpha$ the angle between  $n$ and $\hat{z}$,
i.e. $-\hat{z}_{i} n^{i}=\cos \alpha$. Note, moreover, that
$\delta \chi / \chi$ is the redshift $z(q,q')$ between $q$ and
$q'$. We have therefore
\begin{equation}
\delta \tau=D+\tau z(q,q')-\psi D \cos \alpha
\end{equation}
If we consider a light beam sent from $q'$ to $q$ the last two
terms change sign
\begin{equation}
\delta \tau'=D-\tau z(q,q')+\psi D \cos \alpha
\end{equation}
In order to relate the proper time of clocks to $C$-Killing time,
we first define the proper setting of clocks\footnote{Note that
properly set clocks are not in general synchronized if $\chi$
varies in space.} \\

{\bf Def}. Two neighboring clocks placed in $q$ and $q'$ have been
{\em properly set} according to $C$-simultaneity if the proper
time it takes light to go from $q$ to $q'$ equals the proper time
it takes light to go from $q'$ to $q$ plus a quantity $2\tau
z(q,q')-2\psi D \cos \alpha$ where $D$ is the radar distance
between the two points, $z(q,q')$ is the redshift from $q$ to
$q'$, $\psi$ is the observable scalar and $\alpha$ is the
angle between the direction $qq'$ and the observable $n$. \\

By ``proper time" we mean here the difference of proper times at
the events of arrival and departure of the light beam as measured
by the local clocks. The above definition is automatically
transitive in the hypothesis of small holonomy because of Eq.
(\ref{asdf}).
 For $\bar{C}$-simultaneity it reads \\

{\bf Def}. Two neighboring clocks placed in $q$ and $q'$ have been
{\em properly set} according to $\bar{C}$-simultaneity if the
proper time it takes light to go from $q$ to $q'$ equals the
proper time it takes light to go from $q'$ to $q$ plus a quantity
$2\tau z(q,q')-2\bar{\psi}^{m} D \cos \alpha$ where $D$ is the
radar distance between the two points, $z(q,q')$ is the redshift
from $q$ to $q'$, $\bar{\psi}^{m}$ is the observable given by Eq.
(\ref{magic}) and $\alpha$ is the angle between the direction
$qq'$ and $m$.\\

If the clocks are properly set then $C$-Killing time of an event
in $q$'s worldline is related to the proper time of a clock placed
at $q$ by the simple relation $t=\tau/(1+z(p,q))$. Changing $p$
changes $t$ only by a global factor. Note that the hypersurfaces
of $C$-simultaneity are given by $t=const.$ and not by
$\tau=const.$, moreover note that the term $2\tau z(q,q')$
vanishes if $q$ and $q'$ lie on the same equipotential slice as it
is the case at the surface of a spinning planet.

\section{Conclusions}
In this paper we have introduced the concept of local simultaneity
convention and have began the study of conventions alternative to
the one by Einstein and Poincar\'e. We have focused the study to
conventions whose curvature is proportional to the Riemann tensor,
as they are expected to be useful in the weak field limit.
Remarkably, the simultaneity convention was almost completely
determined by imposing the condition that $\omega_{\alpha}$ should
be timelike and well defined in a sufficiently large domain.
Indeed, $\psi^{m}$ was completely determined, while $\psi^{a}$ and
$\psi^{w}$ were constrained. It then became very natural to
consider the $\bar{C}$-simultaneity convention which was obtained
imposing $\psi^{a}=\psi^{w}=0$. It shares some good feature as its
domain of definition includes events where $m=0$ and reduces to
the Einstein convention in those events. Moreover, it differs only
slightly from the Einstein convention but it is much more well
behaved in flat spacetime or when the Riemann tensor is
particularly weak.

We have seen two examples in flat spacetime where
$\bar{C}$-Killing time have turned out to coincide with the time
coordinate chosen by Letaw and Pfautsch \cite{letaw82} on the
basis of elegance and simplicity.

Unfortunately, it is of no advantage in the most important
application, the synchronization at the surface of a rotating
planet, since the Riemann tensor is not sufficiently weak. In this
respect we note that we are just at the beginning of the study of
local simultaneity conventions. It could be that, in introducing a
non vanishing $\psi^{a}$ some components of the curvature of the
connection could be removed and a new convenient simultaneity
convention could be discovered. We believe that our approach
shares a number of features that makes it preferable over other
approaches:

(a) It is coordinate independent as we always use a transparent
covariant notation. Moreover, it is local and operationally clear
as quantities like the acceleration or the vorticity have a
transparent operational meaning.

(b) It does not rely on Einstein equations or their exact
solutions indeed, contrary to other approaches, we have no need to
assume a special form of the metric in order to apply a local
simultaneity convention, and we do not need to develop a different
method of synchronization for any different spacetime considered.
Moreover, the coordinate time found is universal as the
operational definition of $C$-Killing time makes it possible to
consider $C$-Killing time in different spacetimes.

(c) It does not rely on experimental models of the distribution of
matter (the geoid in the earth case). This is very important since
classical, relativistic or general relativistic effects are
already taken into account in the  local simultaneity convention.
We do not have to change the method of synchronization if we
discover, for instance, that the metric is not the one which was
supposed to be. Once we have proved theoretically that a local
simultaneity convention works, it works independently of the
global information, like the metric, that we may or may not have.

(d) Finally, it does not require particular spacetime symmetries
apart from that of stationarity (local simultaneity conventions
make sense also for non-stationary spacetimes but the study of
this general case appears to be difficult).

In the end we wish to comment on the theorem. We have argued in
the introduction that it would be preferable to obtain a global
simultaneity convention starting from a local simultaneity
convention. The message of the theorem is that globally
well-behaved local simultaneity conventions are not arbitrary but
are severely constrained by the spacetime geometry. Remarkably, it
seems that Nature tells the observers what should be the
convention for the diffusion of time.
 In
principle, global simultaneity conventions are equivalent as they
are related by gauge transformations; however, if one tries to
characterize them operationally through a local simultaneity
convention, only a subset of them survives. In particular, from
our definition of local simultaneity convention it follows that in
flat spacetime and for an inertial frame only the Einstein
simultaneity convention survives since there is only one form
$\omega_{\alpha}$ that can be built using $u_{\alpha}$,
$\eta_{\alpha \beta}$ and $\epsilon_{\alpha \beta \gamma \delta}$
and it is just $\omega_{\alpha}=u_{\alpha}$.

We wonder whether there is some formula which generalizes
$\bar{C}$-simultaneity so as to enlarge the application domain of
local simultaneity conventions. We believe this issue to be worth
studying particularly along the lines introduced in the present
paper.

\section*{Acknowledgments}
I wish to thank the Department of Mathematics of Salamanca, and in
particular Prof. A. L{\'o}pez Almorox for kind hospitality. I also
wish to thank a referee for useful comments. This work was
supported by INFN under the grant $\textrm{n}^{\circ}$ 9503.\\

\section*{Appendix: proof of the theorem}
We collect here  some  useful formulas that will be needed later.
They are just for reference and can be proved straightforwardly
after some calculations. We recall that we are considering a
stationary spacetime.

The covariant derivative of the 4-velocity is
\begin{equation}
u_{\alpha ; \beta}= w_{\alpha \beta}+a_{\alpha} u_{\beta}
\end{equation}
Recalling that $a_{\alpha}=- \ln\chi_{;\alpha}=-k^{\gamma}
k_{\gamma ; \alpha} / \chi^{2}$ the covariant derivative of the
acceleration can be written (note that $ a_{[\alpha;\beta]}=0$)
\begin{eqnarray}
a_{\alpha ; \beta}&=&-\frac{k^{\gamma}_{; \beta} k_{\gamma ;
\alpha}}{\chi^{2}}+\frac{2}{\chi^{4}} k^{\gamma} k_{\gamma ;
\alpha} k^{\delta} k_{\delta ; \beta}+\frac{R_{\beta \delta
\gamma \alpha} k^{\gamma} k^{\delta}}{\chi^{2}} \nonumber \\
&=& a_{\alpha}
a_{\beta}+m_{\beta}u_{\alpha}+m_{\alpha}u_{\beta}+(a^{2}+w^{2})u_{\alpha}u_{\beta}-w^{2}
g_{\alpha \beta}-w_{\alpha} w_{\beta}+R_{\beta \delta \gamma
\alpha} u^{\gamma} u^{\delta} . \label{a}
\end{eqnarray}
The covariant derivative of the vorticity tensor is
\begin{eqnarray}
w_{\alpha \beta ; \gamma}&=&\frac{k_{\alpha ; \beta ;
\gamma}}{\chi}+[u_{\alpha ; \gamma} a_{\beta} + u_{\alpha}
a_{\beta ; \gamma} - u_{\beta ; \gamma} a_{\alpha} - u_{\beta}
a_{\alpha ; \gamma}]-\frac{k_{\alpha ; \beta}}{\chi^{2}} \chi_{;
\gamma} \nonumber \\
&=& -R_{\gamma \delta \alpha \beta} u^{\delta}+ u_{\alpha}
R_{\gamma \delta \eta \beta} u^{\eta} u^{\delta}-u_{\beta}
R_{\gamma \delta \eta \alpha} u^{\eta} u^{\delta} - u_{\alpha}
w_{\delta \gamma} w^{\delta}_{\ \beta}-u_{\alpha} a^{\delta}
w_{\delta \beta} u_{\gamma} + \nonumber \\
&{}& -w_{\gamma \alpha} a_{\beta} -w_{\beta \gamma} a_{\alpha} +
w_{\alpha \beta} a_{\gamma} + u_{\beta} w_{\delta \gamma}
w^{\delta}_{\ \alpha}+ u_{\beta} a^{\delta} w_{\delta \alpha}
u_{\gamma}  , \label{ww}
\end{eqnarray}
while for the vorticity vector we have
\begin{eqnarray}
w^{\eta}_{; \gamma}&=&\frac{1}{2} \epsilon^{\eta \sigma \alpha
\beta} w_{\sigma \gamma} w_{\alpha \beta}+\frac{1}{2}
\epsilon^{\eta \sigma \alpha \beta} a_{\sigma} u_{\gamma}
w_{\alpha \beta}+ w^{\eta} a_{\gamma} - \epsilon^{\beta \sigma
\eta \alpha } a_{\beta} u_{\sigma} w_{\gamma \alpha}-\frac{1}{2}
\epsilon^{\eta \sigma \alpha \beta}u_{\sigma} R_{\gamma \delta
\alpha \beta} u^{\delta} \nonumber \nonumber \\
&=&2w^{\eta} a_{\gamma}-a_{\nu} w^{\nu}\delta^{\eta}_{\gamma}
-\frac{1}{2} \epsilon^{\eta \sigma \alpha \beta}u_{\sigma}
R_{\gamma \delta \alpha \beta} u^{\delta} . \label{w1}
\end{eqnarray}
The covariant derivative of $m_{\alpha}$ is
\begin{eqnarray}
m_{\alpha ; \gamma}&=& w_{\alpha \beta ; \gamma} a^{\beta}+
w_{\alpha \beta} a^{\beta}_{;\gamma} \nonumber \\
&=&-R_{\gamma \delta \alpha \beta} u^{\delta} a^{\beta}+u_{\alpha}
R_{\gamma \delta \eta \beta} u^{\eta} u^{\delta}
a^{\beta}+R_{\gamma \delta \eta \beta} u^{\eta} u^{\delta}
w_{\alpha}^{\ \beta}-u_{\alpha} w_{\delta \gamma}
m^{\delta} + \nonumber \\
&{}& + 2 m_{\alpha} a_{\gamma}+ w_{\alpha \beta} w^{\beta \delta}
w_{\delta \gamma} +w_{\alpha \beta} m^{\beta} u_{\gamma}+w_{\gamma
\alpha} a^{2}+ m_{\gamma} a_{\alpha}  \nonumber \\
 &=& -R_{\gamma \delta \alpha \beta} u^{\delta} a^{\beta}+u_{\alpha} R_{\gamma \delta
\eta \beta} u^{\eta} u^{\delta} a^{\beta}+R_{\gamma \delta \eta
\beta} u^{\eta} u^{\delta} w_{\alpha}^{\ \beta}+u_{\alpha} w_{
\gamma \delta}
m^{\delta} + \nonumber \\
&{}& + u_{\gamma} w_{\alpha \delta} m^{\delta} + 2 m_{\alpha}
a_{\gamma}- w_{\alpha \gamma} (a^{2}+w^{2})+ m_{\gamma} a_{\alpha}
\label{m} ,
\end{eqnarray}
where we have used $w_{\alpha \beta} w^{\beta \delta} w_{\delta
\gamma}=-w^{2} w_{\alpha \gamma}$. Finally, from Eqs. (\ref{a}),
(\ref{ww}), (\ref{w1}), (\ref{m}), we write down the covariant
derivatives of the most relevant scalars
\begin{eqnarray}
(w^{2})_{; \gamma}&=& 2 w^{2} a_{\gamma} - 2 w_{\gamma \alpha}
m^{\alpha}-R_{\gamma \delta \alpha \beta} u^{\delta} w^{\alpha
\beta} , \label{w^2}
\\
(a^{2})_{; \gamma}&=& 2a^{2}a_{\gamma}-2  w_{\gamma \alpha}
m^{\alpha}-2 R_{\gamma \delta \alpha \beta} u^{\delta} u^{\alpha}
a^{\beta} ,  \label{a^2} \\
(m^{2})_{; \gamma}&=&4 m^{2}a_{\gamma}-2(a^{2}+w^{2})w_{\gamma
\alpha} m^{\alpha}+ 2 R_{\gamma \delta \alpha \beta} u^{\delta}
m^{\alpha} a^{\beta}+2R_{\gamma \delta \eta \beta} u^{\eta}
u^{\delta} w^{\beta}_{\ \alpha} m^{\alpha} ,  \label{m^2} \\
(a_{\delta}w^{\delta})_{; \gamma}&=&2(a_{\delta}w^{\delta})
a_{\gamma}+R_{\gamma \delta \beta
\alpha}u^{\beta}u^{\delta}w^{\alpha}-\frac{1}{2} \epsilon^{\eta
\sigma \alpha \beta}a_{\eta}u_{\sigma} R_{\gamma \delta \alpha
\beta} u^{\delta}.
\end{eqnarray}

Since $\{u,w,a,m\}$ is a base (not an orthogonal one) on those
events where $m \ne 0$, any simultaneity convention can be written
in the same events in the form
\begin{equation} \label{conne}
\omega_{\alpha}=u_{\alpha}+\psi^{m} m_{\alpha}+\psi^{a}
a_{\alpha}+ \psi^{w} w_{\alpha} ,
\end{equation}
where the fields $\psi$ depend on local measurable quantities. We
restrict ourselves to fields $\psi$ which depend on $w$, $a$ and
$\theta$. Thus, we do not consider simultaneity conventions
dependent on the Riemann tensor. We split part (i) of the theorem
in two lemmas.

{\bf Lemma (a)}.  In a stationary spacetime let $U$ be the open
set $U=\{x: m(x)> 0 \ \textrm{and} \ a(x) \ne w(x) \}$. Let the
functions $\psi^{m},\psi^{a},\psi^{w}$, depend only on $a$, $w$
and $\theta$. Let us introduce the invertible transformation
\begin{eqnarray}
x_{1}&=&-a_{\delta}w^{\delta}=aw\cos \theta  , \label{trasf1}\\
x_{2}&=&a^{2}+w^{2}   ,  \label{trasf2}  \\
x_{3}&=&a^{2}-w^{2} , \label{trasf3}
\end{eqnarray}
and consider the functions $\psi^{m},\psi^{a},\psi^{w}$, as
dependent on the variables $x_{1}$, $x_{2}$ $x_{3}$. Assume that
the functions $\psi^{m},\psi^{a},\psi^{w}$ are $C^{1}$. Then the
curvature of (\ref{conn}) is proportional to the Riemann tensor in
$U$, regardless of the stationary spacetime considered, if and
only if
\begin{eqnarray}
(\frac{x_{2}^{2}-x_{3}^{2}}{2}-2x_{1}^{2})\frac{\p \psi^{m}}{\p x_{2}}&=& 1-x_{2}\psi^{m} , \label{sec0}\\
\psi^{w}+({x_{2}-x_{3}}) \frac{\p \psi^{w}}{\p x_{2}}&=& -2x_{1}
\, \frac{\p \psi^{a}}{\p x_{2}} . \label{sec}
\end{eqnarray} \\

{\bf Proof}. We introduce the vector valued symmetric operator
which acts on pairs of fields $b$, $c$, given by
\begin{equation}
\langle b,c \rangle^{\eta} =\frac{1}{4}h^{\eta}_{\nu}
\varepsilon^{\nu \beta \alpha \gamma} (b_{\beta} c_{\alpha ;
\gamma}+c_{\beta} b_{\alpha ; \gamma}) ,
\end{equation}
then, omitting the indices,
\begin{equation}
v^{\eta}=\langle u+\psi^{m} m+\psi^{a} a+ \psi^{w} w , u+\psi^{m}
m+\psi^{a} a+ \psi^{w} w \rangle^{\eta}.
\end{equation}
Let us consider the different pairings. The covariant derivative
of $a_{\delta}w^{\delta}$, $a^{2}$ or $w^{2}$ is proportional to
terms in $w^{\alpha}$ and $a^{\alpha}$ plus terms proportional to
the Riemann tensor. As a consequence, the same property holds for
any function $\psi$ of the variables $\cos \theta$, $a$ and $w$.
This fact, along with the calculation of the covariant derivatives
given above, leads us to table \ref{table}. The result of the
pairing is a linear combination of the fields shown in the last
column, apart from terms proportional to the Riemann tensor. We
are looking for a connection $\omega$ such that the sum of the
different pairings vanishes apart from terms proportional to the
Riemann tensor.

The table shows that the only way to remove the term proportional
to $w$, due to the pairing of $u$ with itself, is to have
$\psi^{m} \ne 0$. However it is a non trivial fact that this can
actually be done without adding a term proportional to $a$.

\begin{table}
    \begin{center}
\begin{tabular}{|c|c|c|}
\hline $ \ b \ $ & $c$ & $ \ \langle b,c \rangle \propto \ $ \\
\hline $u$ & $u$ & $w$ \\
 $u$ & $ \ \psi^{m} m  \ $ & $w$, $a$ \\
$u$ & $\psi^{a} a $ & $m$ \\
$u$ & $\psi^{w} w $ & $m$\\ \hline
\end{tabular}
\end{center}
    \caption{Pairings between different terms. The pairings not shown here are proportional to the Riemann tensor. }
    \label{table}
\end{table}

We shall see below that this happens because $\psi^{m}$ must
satisfy a dimensional constraint. Indeed $u$ is dimensionless
whereas $m$ has dimension $[L^{-1}]^{2}$ (we shall say $d=2$),
thus $\psi^{m}$ must have $d=-2$.

When we have removed $w$ from the curvature we can stop or
consider more complicated local simultaneity conventions adding
terms proportional to $a$ and $w$. In general, we expect to need
both since if one adds a term to the curvature proportional to
$m$, the other must remove it. Let us denote by, $\simeq$,
equivalence up to terms proportional to the Riemann tensor. The
simultaneity convention has a curvature proportional to the
Riemann tensor iff
\begin{eqnarray} \label{sys}
\langle u, u \rangle+2\langle u, \psi^{m} m \rangle& \simeq &0 , \\
\langle u, \psi^{a} a \rangle+\langle u, \psi^{w} w\rangle& \simeq
 &0 . \label{sys2}
\end{eqnarray}

The variables $x_i$ have $d_{i}=2$. Since the transformation given
by (\ref{trasf1}), (\ref{trasf2}), (\ref{trasf3}), is invertible
the functions $\psi$ will be considered as functions of $x_{1}$,
$x_{2}$ and $x_{3}$. The covariant derivatives of these scalars
are
\begin{eqnarray}
x_{1;\gamma}&=&d_{1}x_{1}a_{\gamma}-R_{\gamma \delta \beta
\alpha}u^{\beta}u^{\delta}w^{\alpha}+\frac{1}{2} \epsilon^{\eta
\sigma \alpha \beta}a_{\eta}u_{\sigma} R_{\gamma \delta \alpha
\beta} u^{\delta} ,\\
x_{2;\gamma}&=&d_{2}x_{2}a_{\gamma}-4w_{\gamma \alpha}m^{\alpha}-2
R_{\gamma \delta \alpha \beta} u^{\delta} u^{\alpha}
a^{\beta}-R_{\gamma \delta \alpha \beta} u^{\delta}
w^{\alpha \beta},\\
x_{3;\gamma}&=&d_{3}x_{3}a_{\gamma} -2 R_{\gamma \delta \alpha
\beta} u^{\delta} u^{\alpha} a^{\beta}+R_{\gamma \delta \alpha
\beta} u^{\delta} w^{\alpha \beta}.
\end{eqnarray}

Since the functions $\psi$ depend on dimensional variables they
have to satisfy a scaling relation of the form
\begin{equation}
\psi(\lambda^{d_{1}}x_{1},\lambda^{d_{3}}x_{3},\lambda^{d_{3}}x_{3})=\lambda^{d_{\psi}}
\psi(x_{1},x_{2},x_{3}) .
\end{equation}
In particular, $d_{\psi^{m}}=-2$, $d_{\psi^{a}}=-1$,
$d_{\psi^{w}}=-1$. Differentiating with respect to $\lambda$ we
obtain (Euler's theorem), $\sum_{i=1}^{3}d_{i}
x_{i}{\psi_{;i}}=d_{\psi} \psi$. Using the covariant derivatives
of $x_{i}$ we obtain
\begin{equation}
\psi_{;\gamma}=\sum_{i=1}^{3}\frac{\p \psi}{\p x_{i}} x_{i;\gamma}
\simeq(\sum_{i=1}^{3} \frac{\p \psi}{\p x_{i}} d_{i} x_{i} )
-4\frac{\p \psi}{\p x_{2}} w_{\gamma \alpha} m^{\alpha},
\end{equation}
and finally
\begin{equation}
\psi_{;\gamma}\simeq d_{\psi} \psi a_{\gamma} -4\frac{\p \psi}{\p
x_{2}} w_{\gamma \alpha} m^{\alpha}.
\end{equation}
We are ready to solve the system (\ref{sys}), (\ref{sys2}). We
have
\begin{eqnarray}
4\langle u, u \rangle^{\eta}&=& 4 w^{\eta},\\
4\langle u, \psi^{m} m\rangle^{\eta}&=&
h^{\eta}_{\nu}\varepsilon^{\nu \beta \alpha \gamma}( u_{\beta}
m_{\alpha} \psi^{m}_{;\gamma}+ \psi^{m} u_{\beta} m_{\alpha;
\gamma} + \psi^{m} m_{\beta} a_{\alpha}u_{\gamma}) \simeq -w^{\eta}[2(a^{2}+w^{2})\psi^{m}+4m^{2}\frac{\p \psi^{m}}{\p x_{2}}],    \\
4\langle u, \psi^{a}
a\rangle^{\eta}&=&h^{\eta}_{\nu}\varepsilon^{\nu \beta \alpha
\gamma} u_{\beta} a_{\alpha} \psi^{a}_{;\gamma} \simeq -4
(w^{\delta} a_{\delta})\frac{\p \psi^{a}}{\p x_{2}} m^{\eta},\\
4\langle u, \psi^{w} w\rangle^{\eta}&\simeq&
h^{\eta}_{\nu}\varepsilon^{\nu \beta \alpha \gamma}(u_{\beta}
w_{\alpha} \psi^{w}_{;\gamma}+2\psi^{w} u^{\beta}
w_{\alpha}a_{\gamma}+\psi^{w}w_{\beta} a_{\alpha}
u_{\gamma})=(2\psi^{w}+4 w^{2} \frac{\p \psi^{w}}{\p x_{2}})
m^{\eta},
\end{eqnarray}
thus the curvature is proportional to the Riemann tensor if
\begin{eqnarray}
2m^{2}\frac{\p \psi^{m}}{\p x_{2}}&=& 1-(a^{2}+w^{2})\psi^{m} ,\\
\frac{\psi^{w}}{2}+w^{2} \frac{\p \psi^{w}}{\p x_{2}}&=&
(w^{\delta} a_{\delta})\frac{\p \psi^{a}}{\p x_{2}} ,
\end{eqnarray}
where $m^{2}$, $a^{2}$ $w^{2}$ and $w^{\delta} a_{\delta}$ are
understood as functions of $x_{1}$, $x_{2}$ and $x_{3}$.

Let us shows that these conditions are also necessary. Let
\begin{eqnarray}
H(x_{1},x_{2},x_{3})&=&2m^{2}\frac{\p \psi^{m}}{\p x_{2}}-1+(a^{2}+w^{2})\psi^{m} ,\\
K(x_{1},x_{2},x_{3})&=&\frac{\psi^{w}}{2}+w^{2} \frac{\p
\psi^{w}}{\p x_{2}}-(w^{\delta} a_{\delta})\frac{\p \psi^{a}}{\p
x_{2}},
\end{eqnarray}
and consider a flat spacetime $M$. Since the Riemann tensor is
null, the curvature $v$ vanishes only if, for any $y \in M$
\begin{eqnarray}
H(x_{1}(y),x_{2}(y),x_{3}(y))&=&0,\\
K(x_{1}(y),x_{2}(y),x_{3}(y))&=&0.
\end{eqnarray}
If we are able to show that the Killing vector field and the event
$y$ can be chosen so as to obtain any triple $(x_{1},x_{2},x_{3})$
the proof of (a) is completed. It is convenient to consider on $M$
the pseudocylindrical coordinates
\begin{equation}
\dd s^{2}=(\xi^{2}-r^{2}\Omega^{2})\dd \tau^{2}-\dd \xi^{2} - \dd
r^{2}- r^{2}\dd \phi^{2}-2r^{2}\Omega \dd \phi \dd \tau ,
\end{equation}
and the timelike Killing vector field $k=\p_{\tau}$,
$(\xi^{2}-r^{2}\Omega^{2}>0)$. $\Omega$ is a parameter that
satisfies $\Omega^{2} \ne 1$. After some algebra we obtain
\begin{eqnarray}
x_{1}&=&\frac{\Omega}{\xi^{2}-r^{2}\Omega^{2}}, \\
x_{2}&=& (1+\Omega^{2}) \frac{\xi^{2}+\Omega^{2}r^{2}}{(\xi^{2}-r^{2}\Omega^{2})^{2}} ,  \\
x_{3}&=&\frac{1-\Omega^{2}}{\xi^{2}-r^{2}\Omega^{2}},
\end{eqnarray}
that fortunately can be inverted to give
\begin{eqnarray}
\Omega&=&\frac{-x_{3}+\sqrt{x_{3}^{2}+4x_{1}^{2}}}{2x_{1}} \label{omega},\\
\xi&=&\sqrt{\frac{\beta_{1}+\beta_{2}}{2}}   \label{uno},\\
r&=&\frac{1}{\Omega}\sqrt{\frac{\beta_{1}-\beta_{2}}{2}},
\label{due}
\end{eqnarray}
with
\begin{eqnarray}
\beta_{1}&=& \frac{x_{2}}{2x_{1}^{2}}\frac{\sqrt{x_{3}^{2}+4x_{1}^{2}}-x_{3}}{\sqrt{x_{3}^{2}+4x_{1}^{2}}}, \\
\beta_{2}&=&\frac{\sqrt{x_{3}^{2}+4 x_{1}^{2}}-x_{3}}{2x^{2}_{1}}.
\end{eqnarray}
Thus given $(x_{1},x_{2},x_{3})$ we choose on flat spacetime a
pseudocylindric coordinate system having a $\Omega$ given by Eq.
(\ref{omega}), then in the events characterized by the coordinates
$\xi$ and $r$ given by Eqs. (\ref{uno}), (\ref{due}), the
kinematical quantities $x_{1}$, $x_{2}$ and $x_{3}$ assume the
required values.  Lemma (a) is proved.

{\bf Lemma (b)}. Let the functions $\psi^{m},\psi^{a},\psi^{w}$ be
 $C^{1}$ in $x_{1}$, $x_{2}$ and $x_{3}$. They satisfy Eqs. (\ref{sec0}), (\ref{sec}), and the associated connection (\ref{conne}) is
timelike in $U$ (and hence it is a simultaneity connection in $U$)
if and only if
\begin{equation}
\psi^{m}=\bar{\psi}^{m} \equiv \frac{a^{2}+w^{2}-\sqrt{(a^{2}+w^{2})^{2}-4m^{2}}}{2m^{2}}, \\
\end{equation}
and there is a $C^{1}$ function $b: \mathbb{R} \to \mathbb{R}$ of
the variable $x_{1}/x_{3}=-a_{\delta}w^{\delta}/(a^{2}-w^{2})$
such that for $m>0$ and $a \ne w$ the following inequality holds
\begin{equation}
\frac{\sqrt{(a^{2}+w^{2})^{2}-4m^{2}}-(a^{2}-w^{2})}{2w^{2}} \,
b^{2}+\frac{m^{2}}{w^{2}} \psi^{a 2}+m^{2} \bar{\psi}^{m 2} <1 ,
\end{equation}
and
\begin{equation}
\psi^{w}=\frac{a_{\delta}w^{\delta}}{w^{2}}
\,\psi^{a}+\frac{b}{w^{2}} \{
\frac{\sqrt{(a^{2}+w^{2})^{2}-4m^{2}}-(a^{2}-w^{2})}{2}\}^{1/2}.
\end{equation}

{\bf Proof}. Let us focus on the first differential equation
(\ref{sec0}) and set $\psi^{m}=\phi/m^{2}$. The equation greatly
simplifies to give
\begin{equation}
\frac{\p \phi}{\p x_{2}}=\frac{1}{2} \quad \Rightarrow \quad
\phi=\frac{x_{2}}{2}+f(x_{1},x_{3}) .
\end{equation}
But $2f/x_{3}$ is dimensionless and therefore there is a function
$g$ such that $2f/x_{3}=g({x_{1}}/{x_{3}})$ and
\begin{eqnarray}
\psi^{m}&=&\frac{a^{2}+w^{2}+(a^{2}-w^{2})g(\frac{-a_{\delta}w^{\delta}}{a^{2}-w^{2}})}{2m^{2}}.
\end{eqnarray}
Let us introduce $\rho=w/a$ and $\xi=x_{1}/x_{3}={\rho \cos
\theta}/{(1-\rho^{2})}$, then
\begin{eqnarray}
m\psi^{m}&=&\frac{1+\rho^{2}+(1-\rho^{2})g(\xi)}{2\rho \sin
\theta} \label{hole}.
\end{eqnarray}
Consider the numerator $N(\rho,\theta)$ of $m\psi^{m}$ for
$\theta=0$,
\begin{equation}
N(\rho,0)=1+\rho^{2}+(1-\rho^{2})g(\frac{\rho}{1-\rho^{2}}),
\end{equation}
and assume there is a value $\bar{\rho} \ne 1$ such that
$N(\bar{\rho},0)\ne 0$. By continuity
\begin{equation}
\lim_{\theta \to 0} N(\bar{\rho},\theta)^{2}=N(\bar{\rho},0)^{2}
> 0,
\end{equation}
\begin{equation}
\lim_{\theta \to 0}|m\psi^{m}|(\bar{\rho},\theta)=+ \infty .
\end{equation}
From the hypothesis it follows that the condition $|m\psi^{m}|<1$
is not satisfied in $U$ and therefore we conclude that the
connection is timelike in $U$ only if $\forall \rho \ne 1,
N(\rho,0)=0$. The condition $N(\rho,0)=0$ reads
\begin{equation} \label{g}
g(\frac{\rho}{1-\rho^{2}})=\frac{1+\rho^{2}}{\rho^{2}-1}.
\end{equation}
Let us introduce $\lambda=\rho/(1-\rho^{2})$ and
$s=\textrm{sign}({1-\rho^{2}})=\textrm{sign}(\lambda)$. The
inverse formula is $\rho=(-1+s\sqrt{1+4\lambda^{2}})/(2\lambda)$
and plugging this in Eq. (\ref{g})
\begin{equation}
g(\lambda)=-s\sqrt{1+4\lambda^{2}}.
\end{equation}
Finally,
\begin{equation}
m\psi^{m}=\frac{1+\rho^{2}-\sqrt{(1-\rho^{2})^{2}+4\rho^{2}\cos^{2}\theta
}}{2\rho\sin \theta} ,
\end{equation}
and using $\rho=w/a$ we obtain Eq. (\ref{magic}).

Let us come to Eq. (\ref{sec}). It can be rewritten as
\begin{equation}
\frac{\p}{\p x_{2}}(\frac{x_{2}-x_{3}}{2}\psi^{w}+ x_{1}
\psi^{a})= 0.
\end{equation}
Let $p$ be the quantity between brackets. Its square is
independent of $x_{2}$,
\begin{equation}
p^{2}=w^{2}(w^{2}\psi^{w 2}+a^{2}\psi^{a
2}-2a_{\delta}w^{\delta}\psi^{a}\psi^{w})-m^{2}\psi^{a 2}.
\end{equation}
Let $h^{2}$ be the positive term between brackets. From Eq.
(\ref{generic}) we see that a connection is timelike iff
\begin{equation} \label{inequal2}
\omega^{\alpha}\omega_{\alpha}=1-(m\psi^{m})^{2}-h^{2} > 0.
\end{equation}
Dividing $p^{2}$ by $x_{1}$ we obtain a dimensionless function
independent of $x_{2}$, thus there is a function $r(x_{1}/x_{3})$
such that
\begin{equation}
\frac{p^{2}}{x_{1}}=\frac{w^{2}h^{2}-m^{2}\psi^{a
2}}{x_{1}}=r({x_{1}}/{x_{3}}) ,
\end{equation}
from which it follows that
\begin{eqnarray}
h^{2}&=&\frac{x_{1}r({x_{1}}/{x_{3}})+m^{2}\psi^{a 2}}{w^{2}} =
\frac{\rho \cos \theta \, r(\frac{\rho \cos
\theta}{1-\rho^{2}})+\rho^{2}\sin^{2}\theta \,(a
\psi^{a})^{2}}{\rho^{2}}. \label{asd}
\end{eqnarray}
Now, let $\theta=0$, and introduce again
$\lambda=\rho/(1-\rho^{2})$ .
\begin{equation}
h^{2}(\rho, \theta=0)= \frac{1}{\rho} r(\frac{\rho}{1-\rho^{2}})
\end{equation}
We define $b^{2}(\lambda)=h^{2}(\rho(\lambda),0)$ and find
$r(\lambda)=\rho(\lambda)b^{2}(\lambda)$. Using the formula for
$\rho(\lambda)$ and Eq. (\ref{asd}) we obtain
\begin{equation}
h^{2}=\frac{\sqrt{(1-\rho^{2})^{2}+4\rho^{2} \cos^{2} \theta
}-(1-\rho^{2})}{2\rho^{2}}\,b^{2}(\frac{\rho \cos
\theta}{1-\rho^{2}})+\sin^{2}\theta (a\psi^{a})^{2} .
\end{equation}
Recalling the expression of $p$
\begin{equation}
\psi^{w}=\frac{a_{\delta}w^{\delta}}{w^{2}}\psi^{a} \pm
\frac{|b|}{w^{2}\sqrt{2}}[{\sqrt{(1-\rho^{2})^{2}+4\rho^{2}\cos^{2}\theta}-(1-\rho^{2})}]^{1/2}
,
\end{equation}
and defining the sign of $b$ with $b=\pm |b|$ we obtain Eq.
(\ref{w}), while Eq. (\ref{inequal}) follows from Eq.
(\ref{inequal2}). The proof of lemma (b) is complete. Finally point (i) of the theorem follows easily from lemmas (a) and (b).\\

{\bf Proof of (ii)}. The expression of $\bar{\psi}^{m}$ makes
clearly sense also for those points where $m>0$ but $a=w$. Let
$U'=\{x: m>0, \, x \notin B \}$. In the open set $U'$ the variable
$x=(a^{2}+w^{2})/(2|m|)$ satisfies $x>1$. Moreover we can write in
$U' \subset C$
\begin{equation} \label{kij}
|m\bar{\psi}^{m}|=x-\sqrt{x^{2}-1} .
\end{equation}
Let $p \in C-U'$, $q \in C$ and consider the limit $q \to p$.
Since $a^{2}+w^{2}>0$ and $m=0$ in $p$ we have that $x(q) \to +\
\infty$ as $q \to p$. But in this limit $|m\bar{\psi}^{m}| \sim
1/(2x)$ and hence we can define by continuity
$\bar{\psi}^{m}=1/(a^{2}+w^{2})$ in $C-U'$. Of course this implies
that in these points $\bar{\omega}_{\alpha}=u_{\alpha}$ since
$m^{\alpha}=0$ and hence
$\bar{\omega}_{\alpha}\bar{\omega}^{\alpha}=1$. Moreover,
$\bar{\omega}_{\alpha}\bar{\omega}^{\alpha}>0$ in $U'$ as it
follows from Eq. (\ref{kij}).

The curvature is
\begin{equation}
\bar{v}^{\eta}=\langle u,u \rangle^{\eta}+2 \langle
u,\bar{\psi}^{m}m\rangle^{\eta}+{\bar{\psi}^{m 2}}\langle
m,m\rangle.
\end{equation}
The last term vanishes
\begin{eqnarray}
2\langle m,m\rangle^{\eta}&=&h^{\eta}_{\ \nu}\varepsilon^{\nu
\beta
\alpha \gamma} m_{\beta} m_{\alpha ; \gamma}  \nonumber \\
&=& h^{\eta}_{\ \nu}\varepsilon^{\nu \beta \alpha \gamma}
m_{\beta}(-R_{\gamma \delta \alpha \mu} u^{\delta}
a^{\mu}+u_{\alpha}R_{\gamma \delta \mu
\sigma}u^{\mu}u^{\delta}a^{\sigma}+R_{\gamma \delta \mu
\sigma}u^{\mu} u^{\delta} w^{\ \sigma}_{\alpha}) \nonumber \\
&=& -h^{\eta}_{\ \nu}\varepsilon^{\nu \beta \alpha \gamma}
m_{\beta} R_{\gamma \delta \mu \sigma}  u^{\delta} a^{\sigma}
h^{\mu}_{\ \alpha}=0 .
\end{eqnarray}
By construction the curvature is proportional to the Riemann
tensor. The calculation of the remaining terms is straightforward
but rather lengthy. We give here only a formula, that follows from
Eqs. (\ref{w^2})-(\ref{m^2}), that can help the reader in
recovering the final formula (\ref{dfg}).
\begin{eqnarray}
\bar{\psi}^{m}_{; \gamma}&=&-2a_{\gamma} \bar{\psi}^{m}-4w_{\gamma
\alpha} m^{\alpha} \frac{\p}{\p x_{2}}
\bar{\psi}^{m}(x_{1},x_{2},x_{3}) \nonumber \\
&+&\frac{\bar{\psi}^{m}}{\sqrt{(a^{2}+w^{2})^{2}-4m^{2}}}\{
R_{\gamma \delta \alpha \beta}u^{\delta}w^{\alpha
\beta}+2R_{\gamma \delta \alpha \beta} u^{\delta} u^{\alpha}
a^{\beta} \} \nonumber \\
&+&\frac{2\bar{\psi}^{m 2 }}{\sqrt{(a^{2}+w^{2})^{2}-4m^{2}}}\{
R_{\gamma \delta \alpha \beta} u^{\delta} m^{\alpha} a^{\beta}
+R_{\gamma \delta \eta \beta} u^{\eta} u^{\delta} w^{\beta}_{\
\alpha} m^{\alpha} \} .
\end{eqnarray}


\end{document}